\documentclass{optica-article}

\journal{opticajournal} % for journals or Optica Open

\articletype{Research Article}

\usepackage{lineno}
\usepackage{amsmath}
\usepackage{soul, color, xcolor}
\usepackage{svg}
\usepackage{dcolumn}% Align table columns on decimal point
\usepackage{multirow}
\usepackage{amsfonts}
\usepackage{booktabs}
\usepackage{graphicx}
\usepackage{comment}
\begin{document}

\title{Short-wave infrared broadband up-conversion imaging by using a noncritical phase matched bulk KTiOPO$_4$ crystal}

\author{Xiaohua Wang,\authormark{1,2,3} Zhaoqizhi Han,\authormark{1,2,3} Zheng-He Zhou,\authormark{1,2,3} Jin-Peng Li,\authormark{1,2,3} Bo-Wen Liu,\authormark{1,2,3} He Zhang,\authormark{1,2,3} Yin-Hai Li,\authormark{1,2,3,4} Zhi-Yuan Zhou,\authormark{1,2,3,4,\dag} and Bao-Sen Shi\authormark{1,2,3,\ddag}}

\address{\authormark{1}Laboratory of Quantum Information, University of Science and Technology of China, Hefei 230026, China\\
\authormark{2}CAS Center for Excellence in Quantum Information and Quantum Physics, University of Science and Technology of China, Hefei 230026, China\\
\authormark{3}Anhui Province Key Laboratory of Quantum Network, University of Science and Technology of China, Hefei 230026, China\\
\authormark{4}Anhui Kunteng Quantum Technology Co. Ltd., Hefei 231115, China\\

%\authormark{*}These authors contribute equally to this work.\\
}

\email{\authormark{\dag}zyzhouphy@ustc.edu.cn}
\email{\authormark{\ddag}drshi@ustc.edu.cn}

%% email address is required; see note below about the corresponding author designation

% use {asbstract*} to suppress the copyright line. Copyright information will be added in production

\begin{abstract*} 
Compared to cryogenically cooled conventional detectors, up-conversion detection enables efficient room-temperature short-wave infrared (SWIR) imaging. Although quasi-phase-matching (QPM) in periodically poled crystals offers advantages, the small crystal aperture (typically 1 mm$\times$3 mm) limits resolution. Non-poled crystals enable larger apertures but suffer walk-off aberrations. This work overcomes these limitations by using a noncritical phase matched (NCPM) KTiOPO$_4$ crystal (6 mm$\times$7 mm aperture, 0.5 mm length). Results show resolutions 6$\times$ and 2$\times$ higher than periodically poled crystals in orthogonal directions, with broad conversion band (1.3-2.2 $\mu$m) covering biological and atmospheric windows. The absence of walk-off ensures better image fidelity in up-conversion process. This study presents the first comprehensive characterization of NCPM-based broadband up-conversion imaging, demonstrating performance at the theoretical resolution limit while circumventing drawbacks inherent in alternative up-conversion schemes and conventional detectors.
\end{abstract*}

%%%%%%%%%%%%%%%%%%%%%%%%%%  body  %%%%%%%%%%%%%%%%%%%%%%%%%%
\section{Introduction}
Short-wave infrared (SWIR, 1–2.5 $\mu$m) has emerged as a critical spectral regime across multiple disciplines, where its distinctive optical characteristics drive advanced imaging applications. It operates within atmospheric transmission windows, providing exceptional performance in penetrating haze, fog, and smoke, which is particularly advantageous in remote sensing applications\cite{pawlikowska2017single,li2021single,wu2021non,pisani2012hyperspectral}. The eye-safe nature of SWIR radiation allows for higher illumination power in imaging and light detection and ranging (LIDAR) systems, facilitating more accurate measurements without compromising safety\cite{shang2018adaptive}. In biomedical diagnostics, SWIR light offers superior tissue penetration and reduced scattering compared to shorter wavelengths\cite{zhu20211700,zhu2019noninvasive}, enabling high-contrast imaging of deep structures and improved resolution. Furthermore, SWIR's low solar background and atmospheric absorption at key wavelengths enhance the efficiency of free-space communication systems, supporting high-bandwidth data transmission over long distances\cite{restelli2010improved,liao2017long}. These distinct capabilities establish SWIR as a crucial tool for both scientific research and industrial applications.

Compared to conventional direct detection with specialized infrared detectors, up-conversion imaging offers significant advantages for infrared applications. Critically, it enables room-temperature operation by converting infrared photons into visible light by using nonlinear crystals\cite{dam2012room}, thereby eliminating the need for cryogenic cooling typically required for sensitive SWIR detection. This approach leverages mature, high-performance silicon-based detectors (e.g., CCD/sCMOS) with superior noise characteristics, enabling higher sensitivity and faster frame rates\cite{fang2024wide,ge2023midinfrared}. Furthermore, up-conversion provides access to wavelengths beyond the practical limits of cost-effective photodetector materials (e.g., >1.7 $\mu$m for InGaAs), overcoming fundamental material constraints.

Current up-conversion imaging systems based on bulk crystals face significant trade-offs. Via quasi-phase-matching, single-period poled crystals enable high conversion efficiency\cite{liu2019up,ge2024quantum}, while chirped poled crystals afford large field-of-view and wide bandwidth.\cite{ge2023midinfrared,huang2022wide} However, they are inherently limited by small apertures (typically 1 mm×3 mm), leading to compromised spatial resolution due to the loss of high spatial frequencies on the Fourier plane during the up-conversion process\cite{barh2019parametric}. Conversely, larger-aperture bulk crystals (non-poled) enable higher resolution but generally rely on critical phase matching (CPM), which restricts their angular and spectral bandwidth compared to periodically poled crystals\cite{huang2022wide} and introduces detrimental walk-off effects causing image aberrations\cite{barh2019parametric}. Although previous studies have focused on enhancing up-conversion imaging efficiency and field of view through noncritical phase matching (NCPM)\cite{morishita2000quality,baumert1987noncritically}, they were restricted to wavelengths of up to 1064 nm and did not investigate the concurrent realization of high resolution and broadband capability.

In this work, we address this critical gap by demonstrating high-resolution, broadband SWIR up-conversion imaging based on an NCPM bulk crystal. Using a custom-designed, large-aperture (6 mm×7 mm), short (0.5 mm) KTiOPO$_4$ (KTP) crystal, we achieve broadband up-conversion imaging spanning 1.3 $\mu$m to 2.2 $\mu$m, covering key atmospheric windows and biological transparency bands. Moreover, we demonstrate high-fidelity imaging with a measured finest resolvable linewidth of 12.4 $\mu$m at 1.8 $\mu$m wavelength, 8.8 $\mu$m at 1.3 $\mu$m wavelength, representing estimated 6× and 2× improvements along two orthogonal directions over systems using a typical periodically poled crystal. Walk-off-induced image degradation inherent in CPM schemes is effectively eliminated, confirming the advantage of the NCPM approach. This system provides a viable and cost-effective solution for SWIR imaging, particularly in the challenging 1.7-2.2 $\mu$m region.

\section{Basic principles}
In bulk-crystal-based up-conversion imaging system, according to the Rayleigh criterion, the aperture of the crystal determines a theoretical resolution limit of: 
\begin{equation}
    R_{c} = 1.22\sqrt{2}\lambda f/D_{c}, 
    \label{eq.rc}
\end{equation}
where $D_{c}$ represents the width or the height of the crystal aperture, corresponding to the lateral or the vertical resolution, respectively; $f$ represents the focal length of the imaging lens (L1 in Fig.\ref{fig:setup}); the $\sqrt{2}$ factor originates from the degradation of resolution due to spatial coherence of the illumination beam\cite{dam2012theory}. In this work, we chose the full width at half maximum (FWHM) of the beam intensity as the definition of the beam diameter. The pump beam profile functions as a Gaussian aperture, exhibiting a resolution limit of\cite{barh2019parametric}:

\begin{equation}
    R_{G} = 4\sqrt{\ln(2)}\lambda f/(\pi D_{FWHM}) ,
    \label{eq.rg}
\end{equation}
 where $D_{FWHM}$ is the beam diameter; the $\sqrt{2}$ degradation mentioned above has already been counted in. The overall theoretical system resolution is governed by the larger value between the crystal aperture and the Gaussian aperture. Experimentally, the finest resolvable line width equals half the resolution limit\cite{Pedersen2024long}.

\section{Experimental setup}
\begin{figure}[htbp]
\centering\includegraphics[width=13cm]{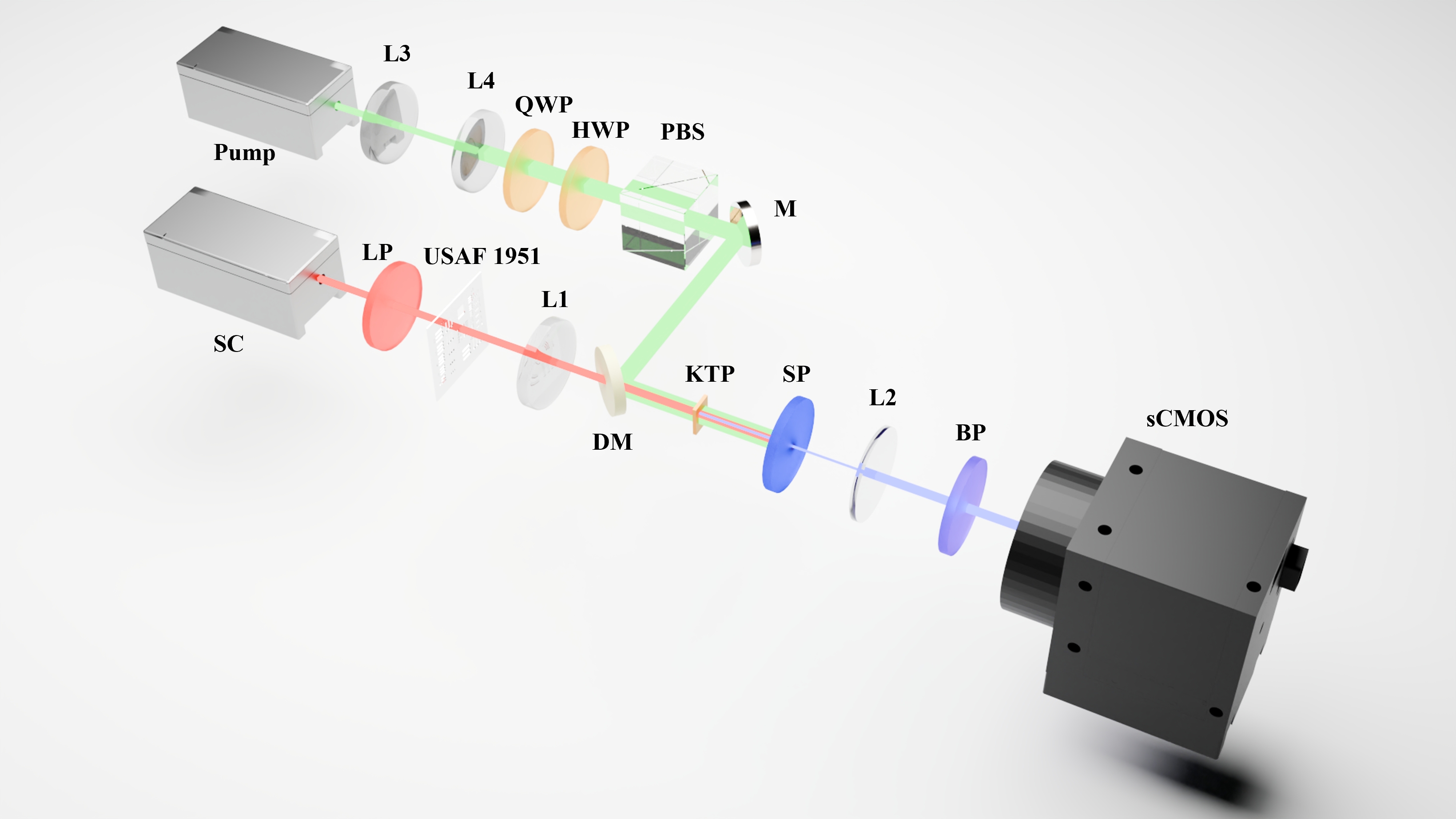}
\caption{Schematic diagram of the experimental setup. SC: supercontinuum pulsed laser; Pump: pump source; L: lens; M: silver mirror; DM: dichroic mirror; LP, SP, BP: long-, short-, and band-pass filters.}
\label{fig:setup}
\end{figure}
The KTP crystal used in our experiment had a length of 0.5 mm and an aperture of 6 mm (width)×7 mm (height). The crystal was cut along its principal axes of refraction to satisfy the type-\uppercase\expandafter{\romannumeral2} ($y+z\rightarrow y$) NCPM condition. During the up-conversion process, the crystal was maintained at room temperature (approximately 24°C), no active temperature control was required. 

The experimental setup is illustrated in Fig.\ref{fig:setup}. The pump source was from a tunable semiconductor laser (KT-TSL-1064, Anhui Kunteng Quantum Technology) coupled into a custom-built ytterbium-doped fiber amplifier, which generated a continuous-wave (CW) pump beam at 1020 nm. The pump beam was expanded by using a two-lens telescope comprising lenses L3 and L4 with focal lengths of 50 mm and 100 mm, respectively, providing a better resolution limit $R_G$ according to Equation (\ref{eq.rg}). The pump beam polarization was aligned parallel to the y-axis of the crystal's refractive index using a wave-plate assembly (QWP and HWP) and polarizing beam splitter (PBS).
The illuminating source in our experiment was a supercontinuum laser (SC-OEM, Wuhan YSL) operating at a repetition rate of 0.5 MHz with a spectral range of 430–2400 nm. Due to its random polarization, no polarization control was required to satisfy the NCPM condition. To mitigate interference from the supercontinuum source on the up-converted signal, we employed sequential long-pass filters with cut-on wavelengths at 1000 nm and 1500 nm. The illumination beam passed through a USAF 1951 resolution target, was focused by a 50-mm focal-length lens (L1), and then combined with the pump beam using a long-pass dichroic mirror (DMLP1180, Thorlabs). This combined beam was subsequently directed into the KTP crystal. Lenses L1 and L2 constituted a 4-f imaging system, with the KTP crystal centered at their common Fourier plane. Featuring a 500-mm focal length, L2 enlarged the up-converted image to ensure each resolution target line spanned multiple camera pixels. Sequential short-pass filters (cut-off wavelength: 1000 nm, 800 nm and 700 nm) rejected residual pump beam, illuminating beam, and background noise. To assess image quality at discrete wavelengths, we introduced band-pass filters (center wavelengths: 578 nm, 650 nm, 690 nm; bandwidth: 10 nm each) to isolate specific spectral bands. The filtered up-converted signal was ultimately captured by an sCMOS camera (Dhyana 95 V2, Tucsen).

\section{Results}
\subsection{Broadband imaging}

\begin{figure}[htbp]
\centering\includegraphics[width=13cm]{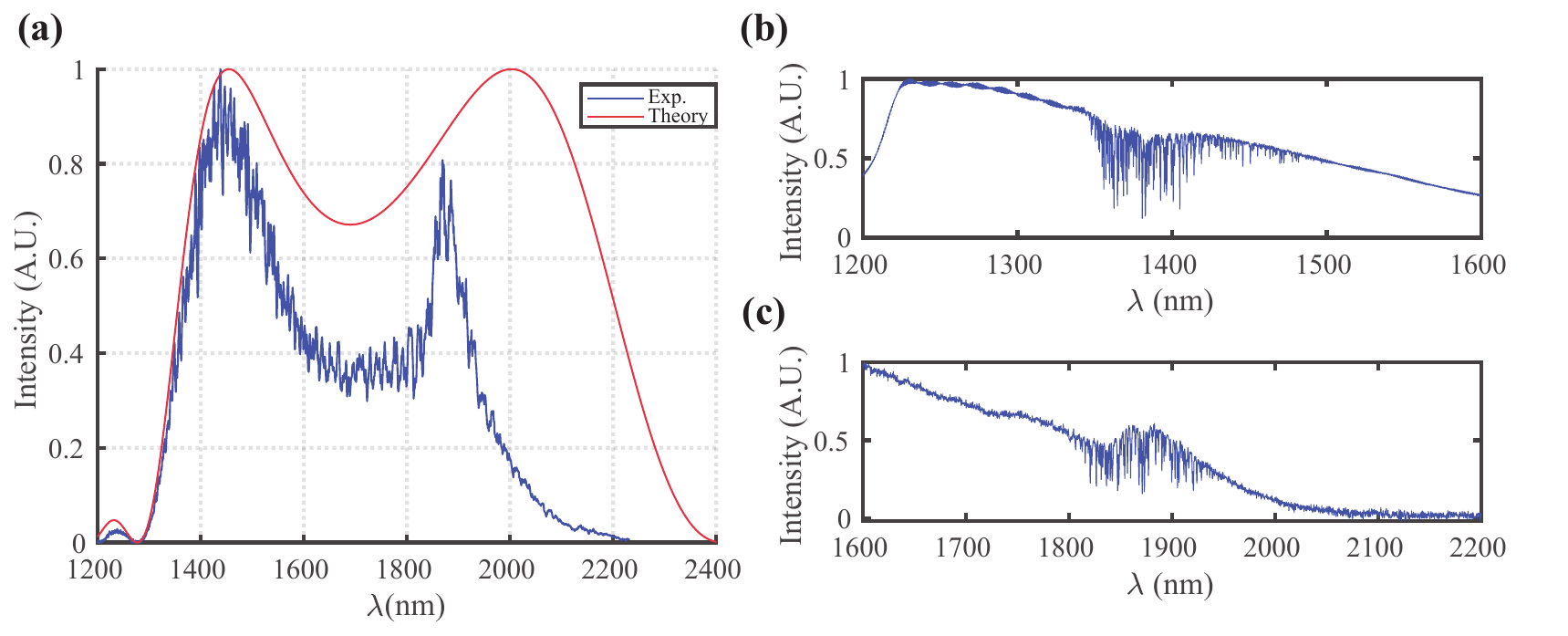}
\caption{Broadband up-conversion performance. (a)  Theoretical and experimental results of up-conversion spectra, with the horizontal axis ticks rescaled to the illumination wavelength based on energy conservation. (b,c) Spectra of illuminating source acquired with two independent spectrometers.}
\label{fig:spectra}
\end{figure}

To characterize the broadband up-conversion performance of the configuration, the up-conversion spectrum was first measured. The up-converted signal was coupled into a single-mode fiber and directed to a spectrometer. The measured spectrum, alongside the theoretical simulation, are shown in Fig.\ref{fig:spectra}(a). Note that the horizontal axis has been converted to the original illumination wavelength based on the principle of energy conservation. The experimental and theoretical results both exhibit consistent spectral profiles, showing two distinct peaks. These peaks and the valley near 1694 nm correspond to the two zero-crossings and an extremum of the phase mismatch, respectively. The discrepancy between experimental and theoretical results at wavelengths longer than the left-side peak primarily stems from the decrease in spectral power density of the supercontinuum laser at longer wavelengths, while the theoretical simulation assumes a constant spectral power density that does not account for this variation.
The spectrum of the supercontinuum source was measured after the KTP crystal using an integrating sphere (Thorlabs, 2P3/M) coupled to a multi-mode fiber. Due to the limited wavelength range of a single spectrometer, the full spectrum of the illumination light (1200–2200 nm) was acquired using two separate spectrometers, as shown in Fig.\ref{fig:spectra}(b) and (c), respectively.
Given the observed decay of spectral power density at longer wavelengths in the supercontinuum source, employing a source with a flatter spectral profile would lead to better agreement between the experimental and theoretical up-conversion spectra specifically at longer wavelengths. Additionally, the central wavelength of the right-side peak in Fig.\ref{fig:spectra}(a) should shift toward longer wavelengths, approaching the theoretical value.

\begin{figure}[htbp]
\centering\includegraphics[width=12cm]{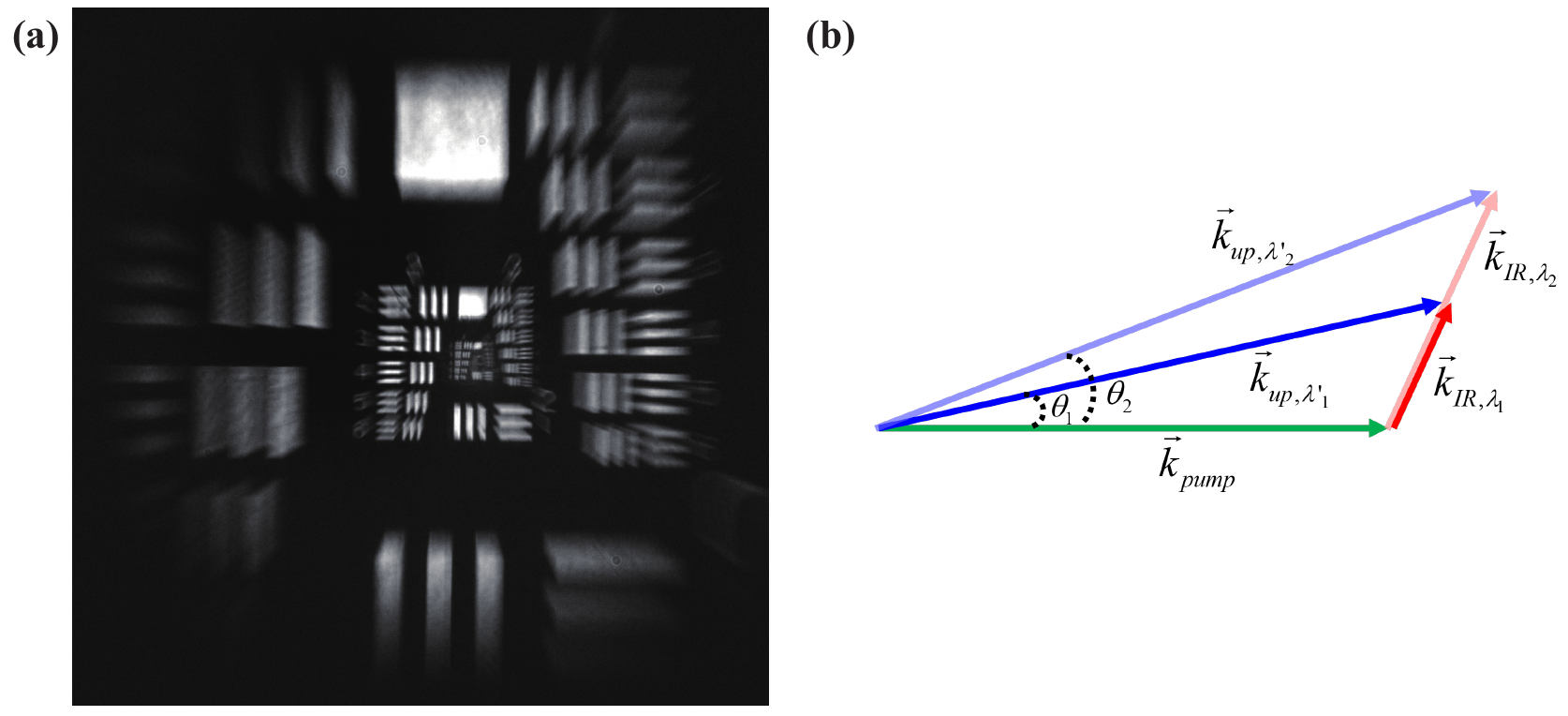}
\caption{Broadband up-conversion imaging. (a) Up-conversion image without employing band-path filters. (b) Illustration of multi-color non-collinear phase matching. The subscripts 'pump', 'IR', and 'up' denote the pump light, infrared light, and up-converted light, respectively.}
\label{fig:blur}
\end{figure}

Fig.\ref{fig:blur}(a) displays the imaging result obtained without using any band-pass filters. Smear-like aberrations surrounding the line-pair patterns are observed, with the smears elongating toward the image center.
This phenomenon primarily reflects the dispersion inherent in the nonlinear process\cite{fang2024wide}. Under non-collinear phase matching conditions, the up-converted light generated from co-propagating beams of different wavelengths is emitted at distinct angles, as illustrated in Fig.\ref{fig:blur}(b). Consequently, the results presented in Fig.\ref{fig:blur}(a) demonstrate the system's capability for broadband imaging.
\subsection{Resolution}
\begin{figure}[htbp]
\centering\includegraphics[width=10cm]{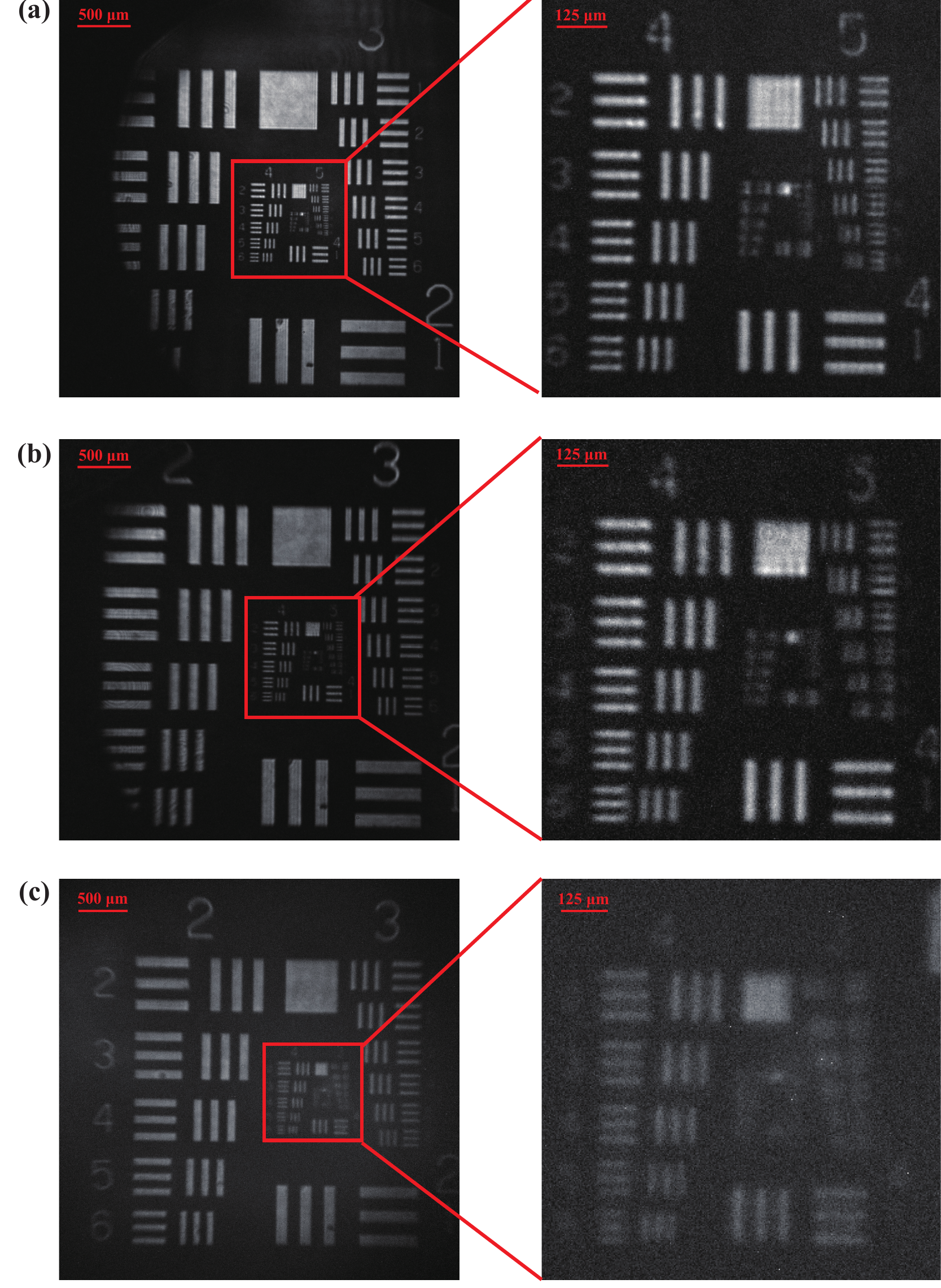}
\caption{Up-conversion images with optimal resolution achieved with band-pass filters (a)578-10, (b)650-10, (c)690-10. The magnified views of the regions enclosed by red boxes in the left column images are displayed in the right column.}
\label{fig:res}
\end{figure}

Subsequently, the band-pass filters specified in the experimental setup were sequentially placed in front of the camera to acquire images at specific visible wavelengths. This selectively transmitted the up-converted light generated by distinct SWIR illumination bands (corresponding to different biological transparency windows), enabling characterization of the system's imaging resolution across different illumination wavelengths.
Fig.\ref{fig:res} presents up-converted images of the USAF 1951 resolution target and magnified views of elements from groups 4 and 5. The corresponding SWIR center wavelengths for the three band-pass filters, along with both the experimentally measured and theoretically calculated resolutions, characterized by the finest resolvable linewidth, are summarized in Table \ref{tab}. For Fig.\ref{fig:res}(a) and (b), the pump beam diameter (FWHM) was 5.3 mm. At a given wavelength, the resolution limit determined by the crystal aperture was poorer than that of the pump beam, according to Equation (\ref{eq.rc}) and (\ref{eq.rg}). Therefore, the theoretical resolution limit was primarily determined by the crystal aperture, not the pump light diameter. The agreement between experimental and theoretical results in Fig.\ref{fig:res}(a) and (b) demonstrates that the system has been fully optimized to exhaust the high-resolution imaging capabilities of this system. For Fig.\ref{fig:res}(c), however, the spectral power density around 2.1 $\mu$m is relatively weak, yielding a low signal-to-noise ratio. To enhance the intensity of the up-converted signal, we reduced the pump beam waist to 3.0 mm in diameter, thereby achieving higher conversion efficiency. Under this pumping condition, the system's resolution limit is determined by the Gaussian aperture of the pump beam, yielding resolution limit of 18.6 $\mu$m according to Equation (\ref{eq.rg}). Experimental results also agree well with theoretical predictions. Future enhancements could enable the 2.1 $\mu$m imaging resolution to approach the theoretical optimum constrained by the crystal aperture, achieved through the implementation of higher-power pump source combined with expanded beam waist.

\begin{table}[htbp]
  \centering
  \caption{Imaging resolution measured using different band-pass filters}
  \resizebox{\linewidth}{!}{
    \begin{tabular}{cccccccc}
    \toprule
    \multicolumn{1}{c}{\multirow{2}[4]{*}{Figure Number}} & \multicolumn{1}{c}{\multirow{2}[4]{*}{Band-pass Filter}} & \multicolumn{1}{c}{\multirow{2}[4]{*}{Center Wavelength (nm)}} & \multicolumn{1}{c}{\multirow{2}[4]{*}{Bandwidth (nm)}} & \multicolumn{2}{c}{Theoretical results ($\mu$m)} & \multicolumn{2}{c}{Experimental results ($\mu$m)} \\
\cmidrule{5-8}          &       &       &       & \multicolumn{1}{c}{Horizontal} & \multicolumn{1}{c}{Vertical} & \multicolumn{1}{c}{Horizontal} & \multicolumn{1}{c}{Vertical} \\
    \midrule
    Fig.\ref{fig:res}(a) & 578-10 & 1334  & 107   & 9.6   & 8.2   & 11.0  & 8.8  \\
    Fig.\ref{fig:res}(b) & 650-10 & 1792  & 152   & 12.9  & 11.0  & 13.9  & 12.4  \\
    Fig.\ref{fig:res}(c) & 690-10 & 2133  & 191   & 18.6  & 18.6  & 19.7     & 19.7 \\
    \bottomrule
    \end{tabular}%
    }
  \label{tab}%
\end{table}%

\section{Discussion and conclusion}
Through detailed investigation of up-conversion imaging resolution, we fully exploited the large-aperture advantage of non-poled crystals, attaining imaging results approaching the theoretical diffraction limit. This demonstrates that NCPM KTP crystals can overcome the inherent resolution constraints imposed by small apertures in periodically poled crystals while enabling broadband imaging. In addition, at a pump power of 800 mW and an illumination power of 514 mW, the system achieved an external quantum efficiency of $1.04\times 10^{-8}$, an imaging signal-to-noise ratio of 12.9, and a field of view of $6.9^\circ$.

Compared to other up-conversion imaging works using CPM non-poled crystals, our approach fundamentally avoids image degradation caused by walk-off effects, thus ensuring better resolution. Our work demonstrates a significantly broader conversion band than previous NCPM-based up-conversion imaging studies\cite{morishita2000quality,baumert1987noncritically}, expanding the spectral range from below 1064 nm to 1.3–2.2 $\mu$m. The broadband imaging enables a wider range of potential applications.

In summary, we have achieved broadband up-conversion imaging detection from 1.3 $\mu$m to 2.2 $\mu$m using an NCPM KTP bulk crystal for the first time. This spectral range comprehensively covers atmospheric transmission windows and three biological tissue optical windows\cite{sordillo2014deep}, providing a powerful tool for hyperspectral imaging and biochemical detection applications. Furthermore, by tuning pump laser wavelengths, this NCPM-based up-conversion imaging scheme can be extended to the mid-infrared region around 3 $\mu$m, potentially enabling high-resolution imaging solutions for diverse mid-infrared applications. 

\begin{backmatter}
\bmsection{Funding}
We would like to acknowledge the support from the National Key Research and Development Program of China (2022YFB3903102, 2022YFB3607700), National Natural Science Foundation of China (NSFC)(62435018), Innovation Program for Quantum Science and Technology (2021ZD0301100), USTC Research Funds of the Double First-Class Initiative(YD2030002023), and Research Cooperation Fund of SAST, CASC (SAST2022-075).

\bmsection{Acknowledgment}
The authors acknowledge Hefei University of Technology for providing access to the spectrometer.

\bmsection{Disclosures}
The authors declare no conflicts of interest.

\bmsection{Data Availability Statement}
Data underlying the results presented in this paper are not publicly available at this time but may be obtained from the authors upon reasonable request.

\end{backmatter}

\bibliography{main}

\end{document}